\documentclass[aps,prl,twocolumn, groupedaddress, showpacs]{revtex4}
\usepackage{amsmath}
\usepackage{amssymb}
\usepackage{graphicx}
\usepackage{color}
\usepackage{tikz}
\usepackage{pgffor}
\usepackage{verbatim}
\begin{document}

\title{Chirality, causality, and fluctuation-dissipation theorems in non-equilibrium steady states}

\author{Chenjie~Wang }
\altaffiliation{Present address: Condensed Matter Theory Center, University of Maryland, College Park, Maryland 20742, USA.}
\affiliation{Department of Physics, Brown University, Providence, Rhode Island 02912, USA}

\author{D.~E.~Feldman}
\affiliation{Department of Physics, Brown University, Providence, Rhode Island 02912, USA}

\date{\today}

\begin{abstract}

Edges of some quantum Hall liquids and a number of other systems exhibit chiral transport: excitations can propagate in one direction only, e.g., clockwise.
We derive a family of fluctuation-dissipation relations in non-equilibrium steady states of such chiral systems. The theorems connect nonlinear response with fluctuations far from thermal equilibrium and hold only in case of chiral transport. They can be used to test chiral or non-chiral character of the system.

\end{abstract}


\maketitle

\begin{figure}
\includegraphics[width=3in]{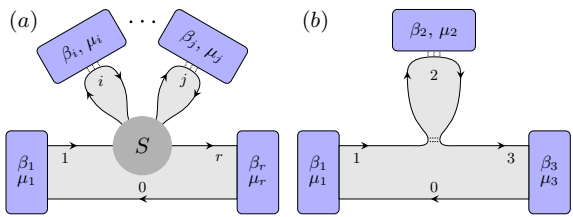}
\caption{(color online) (a) $r$ reservoirs are connected to subsystem S by chiral edge channels $0,1,\dots,r$. The propagation direction along each edge channel is shown with an arrow. Each reservoir is at equilibrium with its own temperature and chemical potential. Panel (b) illustrates a quantum Hall bar. Transport occurs along chiral edges and in quantum point contacts. Dotted lines show tunneling between edges in the  point contacts.}
\end{figure}

According to the causality principle, past events influence the future but the future  has no effect on the past. This principle has no general counterpart in terms of the spatial separation of events: consequences of some events can be felt in every point after a sufficient wait time. A spatial version of the causality principle emerges in low-energy effective theories of some many-body systems. The best known example is the integer quantum Hall effect (QHE): low-energy excitations are confined to the edges and can propagate only clockwise or counterclockwise \cite{wen}. This can lead to a situation in which earlier events affect only those future events that occur ``downstream''. Similar chiral transport is possible in a number of other systems: some fractional quantum Hall liquids \cite{wen}, interfaces of topological insulators, superconductors and ferromagnets \cite{fu-kane,kt}, surface states in 3D QHE
and so on. The simplest example comes from the statistical mechanics models of traffic \cite{helbig}:  chiral transport is possible on a network of one-way roads as long as no traffic jams form.

In this paper we explore consequences of the extended causality principle in chiral systems. Causality is crucial for linear response theory. One of its celebrated results is the fluctuation-dissipation theorem (FDT). We show that a family of generalized FDTs holds in chiral systems. While the usual FDT applies in thermal equilibrium only, our theorems are also valid in non-equilibrium steady states.

The simplest relation \cite{kf,FLS,FS} of such sort was derived for the exactly-solvable chiral Luttinger liquid model with a single impurity. We have recently found an FDT-type relation between the current noise and nonlinear conductance in a general chiral system in a non-equilibrium steady state in a three-terminal geometry
\cite{wang-feldman}. In this paper we prove a much more general result: we express nonlinear responses of the currents of various conserved quantities,
such as the electric current and thermal current, in terms of the second and higher order cumulants of the statistical distributions of the currents in a non-equilibrium steady state in a multi-terminal system with an arbitrary number of terminals. The generalization is achieved due to a much simpler approach.
The result for the chiral Luttinger liquid model follows from its technically difficult exact solution \cite{FS}. A more general result \cite{wang-feldman} was obtained with a simpler but still rather subtle method, generalizing the equilibrium Kubo formalism.  In this paper we use a completely different trick based on fluctuation relations  \cite{ft1,ft2}.

The fluctuation theorem has been used in Ref. \onlinecite{saito08} to derive universal relations for nonlinear transport coefficients in the absence of time-reversal symmetry (see also Refs. \cite{saito09,sb,fb,nasb10,lopez,nakamura10,nakamura11,safi}). The results of Ref. \cite{saito08} hold irrespective of chirality.
Our results apply to chiral systems only and can thus be used to test transport chirality experimentally. This problem is of great interest \cite{feldman-li,granger,bid,deviatov,yacoby} for the
QHE physics, in particular, at the filling factor $5/2$. Indeed, the question of chirality of the edge transport is relevant for the current search for non-Abelian anyons at that filling factor \cite{feldman-li,bid}.
Even in better understood QHE states, such as the Laughlin series, there is no complete theory of edge transport. In particular, the chiral Luttinger liquid model \cite{wen} faces difficulties
(for a review of recent experiments see Ref. \onlinecite{review-heiblum}). Our theorems provide a way to test its basic assumption of chiral edge transport \cite{wang-feldman, deviatov}
and touch upon the problem of nonlinear transport in QHE \cite{molenkamp,siddiki}.

Non-equilibrium fluctuation-dissipation relations have been derived in various classical systems \cite{cl1,cl2,cl3,cl4,cl5,cl6,cl7,cl8,cl9,cl10,cl11,cl12}. In contrast to Refs.
\onlinecite{cl1,cl2,cl3,cl4,cl5,cl6,cl7,cl8,cl9,cl10,cl11,cl12}, we consider chiral transport and focus on topological quantum systems in some of which
chiral transport has received experimental support \cite{granger,bid}.

We begin with a proof of a relation between the nonlinear conductance and the current noise (second order current correlation function) and then generalize it for higher order correlation functions.

 The system (Fig.~1) consists of a central subsystem S and $r$ reservoirs which we number with $k=1,\dots,r$. We assume that the reservoirs are connected to S by chiral edge channels as shown in Fig. 1.
Subsystem S may but does not have to be chiral (see an illustration in Fig. 1b).
There is no bulk transport outside system S and reservoirs. Chiral edges 0 and $r$ are fully absorbed by reservoirs 1 and $r$ respectively. In the case of a QHE system, this means that the resistance of the reservoirs is much lower than the Hall resistance of the QHE bar. We do not make assumptions about the strength of the interaction between the remaining reservoirs $2,\dots,r-1$ and chiral edges.
We assume that transport along edge 0 is uncorrelated with what happens on the other chiral edges in Fig. 1, {\it i. e.}, all forces are short-ranged. For a QHE bar this implies screening of the long-range Coulomb interaction by a gate whose distance from the 2D electron gas is shorter than the distance between edge 0 and the other edges. Edge 0 must also be far enough from the other edges to prevent transport into them through the bulk from edge 0 \cite{fs,gg}. The absence of such leakage current can be verified by testing the quantization of the Hall conductance after all side reservoirs $2,\dots, r-1$ are disconnected.
Each reservoir is maintained at its temperature $T_k=1/\beta_k$ and chemical potential $\mu_k$. Unless all temperatures and all chemical potentials are the same, the system is in a nonequilibrium steady state.
In order for the chiral description to hold, $T_k$  and $\mu_k$ must be lower than the QHE gap.
Let the electric current $I_i$ and heat current $J_i$ flow into reservoir $i$. Our main result is that the zero-frequency cross-noises $ S_{1i}=\lim_{\omega\rightarrow 0}\langle I_1(\omega)I_i(-\omega)\rangle$ and
$S^h_{1i}=\lim_{\omega\rightarrow 0}\langle J_1(\omega)J_i(-\omega)\rangle$,  $i\neq 1, r$, are related to the nonlinear responses of $I_i$ and $J_i$ to the electrostatic potential $V_1$ and temperature $T_1$:
\begin{equation}
\label{equation1}
 S_{1i}=-T_1 \frac{\partial I_i}{\partial V_1},
\end{equation}
\begin{equation}
\label{equation2}
S^h_{1i}=-(T_1)^2\frac{\partial J_i}{\partial T_1}.
\end{equation}
The Boltzmann constant $k_B$ is set to 1 throughout the paper.

Our method is connected with the approach of Refs. \cite{andrieux09, campisi10}.
We consider the following protocol:
Initially, subsystem S and the reservoirs are decoupled. An interaction $\mathcal V(t)$ that allows particle and energy exchange with the reservoirs is turned on at times $0\le t\le \mathcal{T}$. The interaction is turned off at $t\ge \mathcal{T}$. At $t\le 0$, reservoir $i$ is at equilibrium with an inverse temperature $\beta_i=1/T_i$ and a chemical potential $\mu_i=qV_i$, where $q$ is the charge of a charge carrier and $V_i$ the electric potential. We use only one set of chemical potentials and thus assume that only one carrier type is present.
The initial state of finite subsystem S is irrelevant. It is convenient to regroup S (and the rest of grey area in Fig. 1) with one of the reservoirs\cite{andrieux09}, for example, the $r$-th reservoir. The interaction $\mathcal V(t)$ becomes a constant $\mathcal V_0$ when fully turned on during $\tau\le t\le \mathcal{T}-\tau$. We assume that $\tau\ll\mathcal{T} $ and $\mathcal{T}$ is much longer than the relaxation time so that the system remains in a steady state during most of the time interval $\mathcal{T}$.

The above process is called a {\it forward} process in the formalism of fluctuation relations. We also need to study a {\it backward} process which can be described as a forward process in the time-reversed twin system with the opposite chirality (i.e., the directions of all arrows must be reversed in Fig. 1; see Supplementary material for a detailed discussion).
We assume that the initial temperatures and chemical potentials of all reservoirs are the same in the forward and backward processes.

Consider now the changes $\Delta N_i=N_i(t=\mathcal{T})-N_i(t=0)$ in the particle number $N_i$ in each reservoir and the changes  $\Delta E_i=E_i(t=\mathcal{T})-E_i(t=0)$
in the energy $E_i$ in each reservoir. The total energy and particle number are conserved, so that $\sum_i\Delta E_i=\sum_i\Delta N_i=0$.
We introduce the joint probability distribution of the particle number and energy changes $P[\Delta\mathbf E, \Delta \mathbf N;\pm]$,  where the vector $\Delta\mathbf E=\{\Delta E_i\}$, $\Delta \mathbf N=\{\Delta N_i\}$, and
the ``+'' or ``-'' sign in the last argument of $P$ refers to the forward or backward process respectively. According to the fluctuation relation (a derivation can be found in Ref. \cite{andrieux09}  and Supplementary material)

\begin{equation}
\frac{P[\Delta\mathbf E, \Delta \mathbf N;+]}{P[-\Delta\mathbf E, -\Delta \mathbf N;-]}=\prod_{i} e^{\beta_i(\Delta E_i-\mu_i\Delta N_i)} \label{eq-fr}.
\end{equation}

Given a distribution $P[\Delta\mathbf E, \Delta \mathbf N;\nu]$ with $\nu=\pm$, we are able to calculate the heat currents $ J_i^\nu=\lim_{\mathcal T\rightarrow \infty}\langle(\Delta E_i-\mu_i\Delta N_i)\rangle_\nu/\mathcal T$ and their correlation functions, as well as the electric currents $ I_i^\nu=\lim_{\mathcal T\rightarrow \infty}\langle q   \Delta N_i\rangle_\nu/\mathcal T$ and their correlation functions. The triangular brackets mean taking average with respect to $P[\Delta\mathbf E, \Delta \mathbf N;\nu]$. The limit $\mathcal T\rightarrow\infty$ is taken so that the steady-state quantities are obtained. It is convenient to define a cumulant generating function
\begin{align}
\label{eq-gr}
&Q(\mathbf x,\mathbf y, \boldsymbol{\beta }, \boldsymbol{\mu}; \nu) = \lim_{\mathcal T\rightarrow \infty}\frac{1}{\mathcal T}\ln \bigg\{\int\prod_{i<r}\left( d \Delta E_i d\Delta N_i \right)\nonumber \\&\times e^{ -\sum_i x_i\Delta E_i-\sum_i y_i\Delta N_i}  P[\Delta\mathbf E, \Delta \mathbf N;\boldsymbol{\beta}, \boldsymbol{\mu}; \nu]\bigg\},
\end{align}
where the vectors $\mathbf x = \{x_i\}$, $\mathbf y =\{y_i\}$,  $\boldsymbol \beta =\{\beta_i\}$, and $\boldsymbol \mu =\{\mu_i\}$.
The $\mu$-dependence of $Q$ cannot be reduced to a dependence on the differences $\mu_i-\mu_j$ even in the QHE context because of the screening gate.
Since $P$ is normalized, we have  $Q(\mathbf 0, \mathbf 0, \boldsymbol \beta, \boldsymbol\mu; \nu)=0$.
The currents and their correlation functions can be obtained by taking derivatives of $Q(\mathbf x,\mathbf y,\boldsymbol\beta,\boldsymbol\mu; \nu)$ over $x_i$ or $y_i$ and then setting $\mathbf x=\mathbf y=0$. The $n$-th order heat current correlation functions
\begin{align}
C_{ij\cdots k}^{h(n),\nu} =&(-1)^n  [\partial_{x_i}-\mu_i\partial_{y_i}][\partial_{x_j}-\mu_j\partial_{y_j}]\nonumber\\
&\cdots[\partial_{x_k}-\mu_k\partial_{y_k}]Q(\mathbf 0, \mathbf 0, \boldsymbol\beta, \boldsymbol\mu;\nu)\label{eq-current},
\end{align}
and the $n$-th order electric current correlation functions
\begin{align}
C_{ij\cdots k}^{(n),\nu} = (-q)^n\partial_{y_i}\partial_{y_j}\cdots\partial_{y_k}Q(\mathbf 0, \mathbf 0,\boldsymbol\beta,\boldsymbol\mu;\nu).\label{eq-noise}
\end{align}
At $n=1$, $C_{i}^{h(1),\nu}$ and $C_{i}^{(1),\nu}$ are just the heat current $J_i^\nu$ and the electric current $I_i^\nu$. The second-order correlation functions are the low-frequency noises $S_{ij}^{h,\nu}\equiv C_{ij}^{h(2),\nu}$ and $S_{ij}^{\nu}\equiv C_{ij}^{(2),\nu}$ [this definition differs by a factor of 2 from Ref. \onlinecite{wang-feldman}]. Higher-order correlation functions have also been studied experimentally \cite{3dmoment}.
The fluctuation relation (\ref{eq-fr}) leads to a symmetry of the generating function
\begin{equation}
Q(\mathbf x,\mathbf y,\boldsymbol\beta, \boldsymbol\mu; +)=Q(\boldsymbol\beta-\mathbf x, \boldsymbol\xi- \mathbf y,  \boldsymbol\beta, \boldsymbol\mu; -),\label{eq-sym}
\end{equation}
with $\boldsymbol\xi=\{-\beta_i\mu_i\}$.

As discussed above, edge 0 (Fig. 1) is independent of the remaining edges. In other words, our protocol results in two statistically independent transport processes: charge and energy transfer along the lower edge 0 and along the remaining edges in the upper part of the system. This means that the distribution function can be rewritten as
\begin{align}
P[\Delta\mathbf E, \Delta \mathbf N;\nu] = \int d\Delta E_1' d\Delta N_1'\, P_1[\Delta E_1',\Delta N_1';\nu]P_2[\Delta E_1\nonumber \\
-\Delta E_1',\Delta N_1-\Delta N_1',  \Delta E_2,\dots\Delta E_{r-1},  \Delta N_1,\dots,\Delta N_{r-1};\nu],
\end{align}
where $P_1[\Delta E_1',\Delta N_1';\nu]$ is the probability to transport $\Delta E_1'$ energy and  $\Delta N_1'$ particles into reservoir 1 along the lower edge
(edge 0 in Fig. 1), and $P_2[\Delta E_1'',\Delta N_1'',  \{\Delta E_s\},  \{\Delta N_s\};\nu]$ is the probability to transport $\Delta E_1''$ energy and $\Delta N_1''$ particles along edge 1 and change the energy and particle numbers in reservoirs $2,\dots,r-1$ by $(\Delta E_i,  \Delta  N_i)$, $1<i<r$ [Remember that $\Delta E_r$ and $\Delta N_r$ are not independent variables because of conservation laws].
Negative $\Delta E_1',~\Delta E_1'',~\Delta N_1'$ or $\Delta N_1''$ mean the energy and/or particle loss by reservoir 1. We now discuss the dependences of $P_1$ and $P_2$ on the temperatures $\beta_i$ and chemical potentials $\mu_i$. In the setup with the ``$+$'' chirality, energy and particles on the lower edge flow out of reservoir $r$ and into reservoir 1. Due to the extended causality principle, the distribution of $\Delta E_1',\Delta N_1'$ depends only on $\beta_r$ and $\mu_r$. Meanwhile, reservoir $r$ receives energy and particles from the upper part of the system but does not provide any feedback, so the transport in the upper part does {\it not} depend on $\beta_r$ and $\mu_r$. Hence, $P_1$ only depends on $\beta_r$ and $\mu_r$ while $P_2$ does {\it not} depend on $\beta_r$ and $\mu_r$. In the setup with the opposite ``$-$'' chirality, the same argument shows that  $P_1$ only depends on $\beta_1$ and $\mu_1$  while $P_2$ does {\it not} depend on $\beta_1$ and $\mu_1$.

In terms of the cumulant generating function, Eq.~(8) means that $Q(\mathbf x,\mathbf y,\boldsymbol\beta, \boldsymbol\mu; \nu)=Q_1+Q_2$ splits into two terms $Q_1$ and $Q_2$, corresponding to $P_1$ and $P_2$ respectively:

\begin{eqnarray}
\label{q1}
Q_1(\mathbf {x},\mathbf {y},\boldsymbol\beta,\boldsymbol\mu;\nu)=\lim_{\mathcal T\rightarrow \infty}\frac{1}{\mathcal T}\ln \bigg\{\int d \Delta E_1' d\Delta N_1' & & \nonumber\\
\times e^{ - (x_1-x_r)\Delta E_1'- (y_1-y_r)\Delta N_1'}  P_1[\Delta E_1', \Delta N_1';\boldsymbol{\beta}, \boldsymbol{\mu}; \nu]\bigg\}; & &
\end{eqnarray}
\begin{eqnarray}
\label{q2}
Q_2(\mathbf {x},\mathbf {y},\boldsymbol\beta,\boldsymbol\mu;\nu)=\lim_{\mathcal T\rightarrow \infty}\frac{1}{\mathcal T}\ln \bigg\{\int d \Delta E_1'' d \Delta N_1'' & & \nonumber\\
\prod_{1<i<r}d \Delta E_i d\Delta N_i
\exp( -x_1\Delta E_1''-\sum_{1<i<r} x_i\Delta E_i-x_r\delta E_r) & & \nonumber\\
\times\exp(-y_1\Delta N_1''-\sum_{1<i<r} y_i\Delta N_i-y_r\delta N_r)  & &\nonumber\\
\times P_2[\Delta E_1'',\Delta N_1'',\{\Delta E_{r>i>1}\},
\{\Delta N_{r>i>1}\};\boldsymbol{\beta}, \boldsymbol{\mu}; \nu]\bigg\}, & &
\end{eqnarray}
where we used the relations $\sum\Delta E_i=\sum \Delta N_i=0$ and defined $\delta E_r=-\Delta E_1''-\sum_{1<i<r}\Delta E_i$, $\delta N_r=-\Delta N_1''-\sum_{1<i<r}\Delta N_i$.
The chirality-induced causality means that $Q_1(\nu=+)$ depends only on $\beta_r$ and $\mu_r$  while $Q_2(\nu=+)$ does {\it not} depend on $\beta_r$ and $\mu_r$. $Q_1(\nu=-)$ depends only on $\beta_1$ and $\mu_1$ and $Q_2(\nu=-)$ does {\it not} depend on $\beta_1$ and $\mu_1$.

We are now ready to prove the steady-state FDT for chiral systems. Let us start with the particle transport. We apply the differential operator $\hat D_i = D_{y_i}-T_i D_{\mu_i}$ to both sides of Eq.~(\ref{eq-sym}). $D_{y_i}$ and
$D_{\mu_i}$ stay for full derivatives over the respective variables.  We obtain
\begin{align}
(\partial_{y_i}-T_i\partial_{\mu_i})Q(\mathbf x,\mathbf y,\boldsymbol\beta, \boldsymbol\mu; +) = -T_i \partial_{\mu_i}Q(\boldsymbol\beta-\mathbf x, \boldsymbol\xi- \mathbf y, \boldsymbol\beta, \boldsymbol\mu; -) \label{eq-dev}
\end{align}
We emphasize that the partial derivative with respect to $\mu_i$ on the right hand side is taken at a fixed $\boldsymbol\xi$. The expression $\xi_i=-\beta_i\mu_i$ should be substituted after the differentiation.
This reflects the difference of the operators $D$ and $\partial$.
We now apply the differential operator $\hat D_j$ on the two sides of Eqs. (\ref{eq-dev}) and set $\mathbf x=\mathbf y=0$ at the end. In terms of the correlation functions~(9), one finds
\begin{equation}
T_j\frac{\partial I_i^+}{\partial V_j}+T_i\frac{\partial  I_j^+}{\partial V_i} = -\mathcal S_{ij}^+ + q^2 T_iT_j\partial_{\mu_i}\partial_{\mu_j}Q
{(\boldsymbol\beta,\boldsymbol\xi, \boldsymbol\beta, \boldsymbol\mu;-)}.\label{eq-11}
\end{equation}
In order to derive Eq. (\ref{eq-11}) we use the identity $\partial_{\mu_i}\partial_{\mu_j}Q{(\mathbf 0,\mathbf 0,\boldsymbol\beta, \boldsymbol\mu;+)}=0$ which follows from $Q(\mathbf 0,\mathbf 0, \boldsymbol \beta,\boldsymbol \mu;\nu)=0$. Note that the last term in Eq.~(\ref{eq-11}) is defined for a system with the ``$-$'' chirality, whereas all other terms refer to the ``$+$'' chirality.

In a chiral system the last term in Eq. (\ref{eq-11}) is zero at $i=1<j\le r$.
This can be seen by writing the partial derivative $\partial_{\mu_1}\partial_{\mu_j}Q(\mathbf x=\boldsymbol\beta,\mathbf y = \boldsymbol\xi, \boldsymbol\beta, \boldsymbol\mu;-)$ as the sum of the derivatives of $Q_1$ and $Q_2$. In the system with  the ``$-$'' chirality, $Q_1$ depends only on $\mu_1$ while $Q_2$ does not depend on $\mu_1$ before we make the substitution $\mathbf x\rightarrow \boldsymbol \beta$ and $\mathbf y\rightarrow \boldsymbol \xi$. Thus, the partial derivatives $\partial_{\mu_1}\partial_{\mu_j}Q_1$ and $\partial_{\mu_1}\partial_{\mu_j}Q_2$ are both zero. Moreover, if $1<j<r$ then the first term of Eq.~(\ref{eq-11}) $\partial I_1^+/\partial \mu_j$ is also zero since $I_1^+$ only depends on $\beta_{1,r}$ and $\mu_{1,r}$ in the system with the ``+'' chirality. Hence in systems with the ``+'' chirality, Eq. (\ref{eq-11}) simplifies to Eq. (\ref{equation1}). This is our main result:
The cross noise between the currents in reservoirs 1 and $j$ is connected to the response of the current in reservoir $j$ to the voltage in reservoir 1 regardless of the non-equilibrium nature of the system. If $j=r$ then the  term $\partial  I_1/\partial V_r$ is not zero. Instead, it equals the conductance $G$ in the two-terminal setup with only two reservoirs 1 and $r$.
This is the case since $\partial  I_1/V_r$ is solely determined by edge 0 and hence does not change if all reservoirs $2,3,\dots,r-1$ are removed.
Thus, at $j=r$ there is an additional contribution $GT_r$ on the left hand side of Eq. (\ref{equation1}).
 In a QHE system with the filling factor $\nu$, $G=\nu e^2/h$.

Similar results can be obtained for heat currents and noises after one applies the differential operators $\hat D_j^h = D_{x_j}-\mu_j D_{y_j}+ D_{\beta_j}$ and $\hat D_1^h$ on both sides of Eq.~(\ref{eq-sym}). For the systems with the ``$+$'' chirality and $j\neq 1,r$, we derive Eq. (\ref{equation2}) with this trick.
For $j=r$, there is an additional term $\partial J_1/\partial \beta_r=-(T_r)^2\partial J_1/\partial T_r$ on the right hand side of Eq.~(\ref{equation2}).
In QHE, the physical meaning of the additional contribution $\partial J_1/\partial T_r$ is the thermal Hall conductance of a two-terminal Hall bar, equal to $\kappa\pi^2T_r/3h$, where $\kappa$ is universal.

We stress that all above response functions are nonlinear responses in non-equilibrium steady states. The FDTs (\ref{equation1}) and (\ref{equation2}) hold for all chiral systems, as long as $T_i$ and $V_i$ are smaller than the energy gap in the bulk.  For non-chiral systems, such as QHE systems with charged or neutral counter-propagating modes at the edges, the last term in Eq.~(\ref{eq-11}) is nonzero in non-equilibrium states and the theorem does not apply. One microscopic mechanism of FDT breaking in non-chiral systems involves energy transport by `upstream' modes from region S (Fig. 1) to reservoir 1 along edge 1. Local heating at the hot spot, where edge 1 enters reservoir 1,
affects the noise of the current, emitted from reservoir 1, as well as the cross-noises $S_{1i}$.

Eqs. (\ref{equation1},\ref{equation2}) differ from the related results of Refs \onlinecite{FS,wang-feldman}. In the case of Ref. \onlinecite{FS}  this reflects a different geometry.
Ref. \onlinecite{wang-feldman} considers a version of Fig. 1 with $r=3$ and $T_r=T_1$ (Fig. 1b). We verify below that the result of
Ref. \onlinecite{wang-feldman} can be derived from Eq. (\ref{equation1}) with $r=3$. Charge conservation implies that for low-frequency components of the electric currents $I_3=-I_1-I_2$. Hence, $S_{33} = S_{11}+S_{22}+2S_{12}$, where $S_{ii}$ denotes the auto-correlation noise and $S_{12}$ is the cross-noise (\ref{equation1}). Since the edges, connected to reservoir 1, are always in equilibrium with the temperature $T_1=T_r$, we find $S_{11}=2GT_1$ from the
equilibrium Nyquist formula [remember a missing factor of 2 in our definition of the noise]. Then from the chiral-system FDT (\ref{equation1}) we get
\begin{equation}
S_{33} = S_{22} -2T_1\frac{\partial I_2}{\partial V_1} + 2GT_1,
\end{equation}
in agreement with Ref.~\onlinecite{wang-feldman}. Certainly, the results of the present work go well beyond Refs. \onlinecite{FS,wang-feldman}: we cover multi-terminal geometries,
thermal currents and a family of FDT's for higher-order correlation functions which we derive below.

The strategy of their derivation is the same as above.
We apply $\hat D_i$ on both sides of Eq.~(\ref{eq-sym}) $m$ times and obtain
\begin{eqnarray}
\label{many-cor}
\hat D_i \cdots \hat D_j  \hat D_{k} Q(\mathbf 0,\mathbf 0, \boldsymbol\beta, \boldsymbol\mu;+)= & &\nonumber \\
(-1)^mT_iT_j\cdots T_k\partial_{\mu_i}\cdots\partial_{\mu_j}\partial_{\mu_k} Q(\boldsymbol\beta,\boldsymbol\xi,\boldsymbol\beta, \boldsymbol\mu; -). & &
\end{eqnarray}
We set $m\ge2$, $k=1$ and $j\neq1,r$. The right hand side vanishes due to chirality. The left hand side expresses as a combination of derivatives of correlation functions. It simplifies dramatically
after we combine Eqs. (\ref{many-cor}) for all $m\le n$. Specifically, one can prove by induction the following FDT:
\begin{equation}
\label{last1}
C^{(n),+}_{i\cdots j1} + T_j\frac{\partial  C^{(n-1),+}_{i\cdots \hat j 1}}{\partial V_j} + T_1\frac{\partial  C^{(n-1),+}_{i\cdots j\hat 1}}{\partial V_1} + T_jT_1\frac{\partial^2  C^{(n-2),+}_{i\cdots\hat j\hat 1}}{\partial V_j\partial V_1}=0,
\end{equation}
where a hat above an index  means that the particular index is absent in the index set (but does not mean that its value is absent in the set since we do not exclude a situation with two or more identical indexes). To derive relations between heat current correlation functions and their response functions, we apply the differential operator $\hat D^h_i$ multiple times on both sides of Eq.~(\ref{eq-sym}). We obtain
\begin{equation}
\label{last2}
 C^{h(n)}_{i\cdots j1} + (T_j)^2\frac{\partial  C^{h(n-1)}_{i\cdots \hat j 1}}{\partial T_j} + (T_1)^2\frac{\partial  C^{h(n-1)}_{i\cdots j\hat 1}}{\partial T_1} + (T_jT_1)^2\frac{\partial^2  C^{h(n-2)}_{i\cdots\hat j\hat 1}}{\partial T_j\partial T_1}=0.
\end{equation}
Eqs. (\ref{equation1},\ref{equation2}) are special cases of (\ref{last1}) and (\ref{last2}).

In conclusion, we established a family of fluctuation-dissipation theorems for charge and heat transport in chiral systems in nonequilibrium steady states. The argument combines the formalism of fluctuation relations with the extended causality from chirality. Our results do not hold in nonchiral systems away from equilibrium and can thus be used to probe the chiral character of charge and energy transport. This question is of significant current interest \cite{feldman-li,granger, bid,deviatov,yacoby} in the QHE physics and is relevant in other fields, such as transport in heterostructures of topological insulators.

We acknowledge a helpful discussion with M. Levin. The work was supported by NSF under Grants No. DMR-0544116 and DMR-1205715.

\newpage
\widetext
\appendix

\section{Supplementary material for Chirality, causality, and fluctuation-dissipation theorems in non-equilibrium steady states: Derivation of the fluctuation relation}

To make the paper self-contained we include a brief derivation of the fluctuation relation (\ref{eq-fr}) [Eq. (3) of the main text].
Additional details can be found in Refs. [42,43] and references therein.

The main text defines forward and backward processes. We want to connect the joint distribution functions $P[\Delta\mathbf E, \Delta \mathbf N;\pm]$ of the energy and particle number changes in the reservoirs in the forward and backward
processes.
We first derive an expression for the distribution function in the forward process.

We want to find the statistical distribution of the changes $\Delta N_i=N_i(t=\mathcal{T})-N_i(t=0)$ in the particle number $N_i$ in each reservoir.
Let $\mathcal H_i$  and $\mathcal N_i$ be the Hamiltonian and particle number operators of the $i$-th reservoir ($\mathcal H_r$ includes system S).
The particle numbers conserve in the absence of $\mathcal V(t)$, i.e., $[\mathcal H_i, \mathcal N_i]=0$. Thus, the initial density matrix factorizes into a product of Gibbs distributions in each reservoir,
$\rho_{n}=\prod_{i} e^{-\beta_i[E_{in}-\mu_i N_{in}]}/Z_0^+$, where $Z_0^+$ is the initial partition function and the index $n$ labels the quantum state $|\psi_n\rangle$ with the reservoir energies $E_{in}$ and particle numbers $N_{in}$.
An initial joint quantum measurement of  $\mathcal H_i$ and $\mathcal N_i$ is performed at $t=0$, so that the quantum state of the system collapses to a common eigenstate $|\psi_n\rangle$ with the probability $\rho_n$.
Then the state $|\psi_n\rangle$ evolves according to the evolution operator $ U(t;+)$ determined by the Hamiltonian $\mathcal H(t;+)=\sum_{i} \mathcal H_i +\mathcal V(t)$.
At $t=\mathcal T$, a second joint measurement is taken, leading to the collapse of the system to the state $|\psi_m\rangle$ with the reservoir energies $E_{im}$ and particle numbers $N_{im}$.
The probability to observe such process is $P[m,n]=|\langle\psi_m|U(\mathcal T;+)|\psi_n\rangle|^2\rho_{n}$.
Hence, the joint distribution function of the energy and particle changes $\Delta E_{i,mn}=E_{im}-E_{in}$ and $\Delta N_{i,mn}=N_{im}-N_{in}$ is
\begin{equation}
%\label{P+}
\tag{A1}
P[\Delta\mathbf E, \Delta \mathbf N;+] = \sum_{mn}\prod_{i}\delta(\Delta E_i - \Delta E_{i,mn})\delta(\Delta N_i - \Delta N_{i,mn}) |\langle\psi_m|U(\mathcal T;+)|\psi_n\rangle|^2\rho_{n} \label{eq-forward},
\end{equation}
where the vector $\Delta\mathbf E=\{\Delta E_i\}$ and $\Delta\mathbf N=\{\Delta N_i\}$. The total energy and particle number are conserved, so that $\sum_i\Delta E_i=\sum_i\Delta N_i=0$.

We also need to study the backward process which can be described as the forward process in the time-reversed twin system with the opposite chirality.
The time evolution operator $U(t;-)$ of the twin system is determined by the Hamiltonian $\mathcal H(t;-)=\Theta \mathcal H(\mathcal T -t;+)\Theta^{-1}$, where $\Theta$ is the time-reversal operator. The $i$-th reservoir has the Hamiltonian $\Theta\mathcal H_i\Theta^{-1}$. Clearly, $\Theta|\psi_m\rangle$ is a common eigenstate of $\Theta\mathcal H_i\Theta^{-1}$ and $\mathcal N_i=\Theta\mathcal N_i\Theta^{-1}$ with the eigenvalues $E_{im}$ and $N_{im}$.
We assume that the initial temperatures and chemical potentials of all reservoirs are the same in the forward and backward processes. Note that such initial temperatures and chemical potentials are not necessarily
the same as the final thermodynamic parameters in the forward process for large but finite reservoirs.
 Performing two quantum measurements at $t=0$ and $t=\mathcal T$ in the beginning and end of the backward process, one finds the distribution of the energy and particle number changes, similar to Eq. (\ref{eq-forward}):
\begin{align}
& P[\Delta\mathbf E, \Delta \mathbf N;-] = \sum_{mn}\prod_{i}\delta(\Delta E_i - \Delta E_{i,nm})\nonumber\\ &\times\delta(\Delta N_i - \Delta N_{i,nm}) |\langle\psi_n|\Theta^{-1}U(\mathcal T;-)\Theta|\psi_m\rangle|^2\rho_{m},\label{eq-backward}
\tag{A2}
\end{align}
where the initial density matrix $\rho_{m}=\prod_{i} e^{-\beta_i[E_{im}-\mu_i N_{im}]}/Z_0^-$.
It follows from the antiunitarity of  $\Theta$ that $Z_0^+=Z_0^-$.

The evolution operators have an important property [41]
\begin{equation}
\Theta^{-1}U(\mathcal T;-)\Theta = U^\dag(\mathcal T; +).\label{eq-U}
\tag{A3}
\end{equation}
Combining Eq.~(\ref{eq-forward}), (\ref{eq-backward}) and (\ref{eq-U}), we recover the fluctuation relation
\begin{equation}
\frac{P[\Delta\mathbf E, \Delta \mathbf N;+]}{P[-\Delta\mathbf E, -\Delta \mathbf N;-]}=\prod_{i} e^{\beta_i(\Delta E_i-\mu_i\Delta N_i)} \label{eq-fr}.
\tag{A4}
\end{equation}

\end{document}